\newif\ifdraft
\newif\iffull
\newif\ifcomment
\newif\iflatexdiff
\newif\ifbibtex
\newif\ifpreprint
\bibtextrue
\preprinttrue
 \fulltrue
\def\dvers{v2.3} 

\newcommand{\papertitle}{Anomalous  evolution of the near-side jet peak shape \\ in Pb--Pb collisions at \snn\ = \unit[2.76]{TeV}}
\newcommand{\paperabstract}{The measurement of two-particle angular correlations is a powerful tool to study jet quenching in a $\pt$ region inaccessible by direct jet identification. In these measurements  pseudorapidity (\Deta) and azimuthal (\Dphi) differences are used to extract the shape of the near-side peak formed by particles associated to a higher $\pt$ trigger particle ($1 <~\ptt <$~\unit[8]{\gevc}). A combined fit of the near-side peak and long-range correlations is applied to the data allowing the extraction of the centrality evolution of the peak shape in Pb--Pb collisions at \snn\ = \unit[2.76]{TeV}. A significant broadening of the peak in the $\Deta$ direction at low $\pt$ is found from peripheral to central collisions, which vanishes above \unit[4]{\gevc}, while in the $\Dphi$ direction the peak is almost independent of centrality. For the 10\% most central collisions and $1 < \pta <$~\unit[2]{\gevc}, $1 < \ptt <$~\unit[3]{\gevc} a novel feature is observed: a depletion develops around the centre of the peak. The results are compared to pp collisions at the same centre of mass energy and to AMPT model simulations. The comparison to the investigated models suggests  that the broadening and the development of the depletion is connected to the strength of radial and longitudinal flow.}

\ifpreprint
\documentclass[ALICE,manyauthors]{cernphprep}
\usepackage[comma,square,numbers,sort&compress]{natbib}
\else
\documentclass[10pt,aps,prl,twocolumn,superscriptaddress,altaffilletter,nobibnotes,nofootinbib]{revtex4-1}
\usepackage{preprintcover}
\PreprintIdNumber{CERN-EP-2016-229}
\PreprintCoverPaperTitle{\papertitle}
\PreprintCoverAbstract{\paperabstract}
\PreprintJournalName{PRC}
\fi
\usepackage{hyperref}
\usepackage{graphicx}  
\usepackage{dcolumn}   
\usepackage{bm}        
\usepackage{amssymb}   
\usepackage{amsfonts}
\usepackage{graphics}
\usepackage{epsfig}
\usepackage{epstopdf}
\usepackage[usenames]{color}
\usepackage{multirow}
\usepackage{units}
\iflatexdiff
\RequirePackage[normalem]{ulem}
\RequirePackage{color}\definecolor{RED}{rgb}{1,0,0}\definecolor{BLUE}{rgb}{0,0,1}

\fi
\ifpreprint
\else

\fi

\newcommand{\gevc}         {GeV/\ensuremath{c}}

\newcommand{\ptt}          {\ensuremath{p_{\mathrm{T, trig}}}}
\newcommand{\pta}          {\ensuremath{p_{\mathrm{T, assoc}}}}

\newcommand{\ITS}          {\rm{ITS}}

\newcommand{\TPC}          {\rm{TPC}}

\newcommand{\pp}           {pp}

\newcommand{\s}            {\ensuremath{\sqrt{s}}}
\newcommand{\pt}           {\ensuremath{p_{\mathrm{T}}}{ }}

\newcommand{\snn}          {\ensuremath{\sqrt{s_{\mathrm{NN}}}}}

\newcommand{\dd}           {\ensuremath{\mathrm{d}}}

\newcommand{\Dphi}         {\ensuremath{\Delta\varphi}}
\newcommand{\Deta}         {\ensuremath{\Delta\eta}}
\newcommand{\Ntrig}        {\ensuremath{N_{\mathrm{trig}}}}

\newcommand{\dNassoc}      {\ensuremath{\frac{\dd^2N_{\mathrm{assoc}}}{\dd\Deta\dd\Dphi}}}

\newcommand{\red}[1]       {\textcolor{red}{#1}}

\newcommand{\warn}[1]      {{\small\textbf{\red{(!}\footnote{\textbf{\red{(!)}}~#1}\red{)}}}\marginpar{\textbf{\red{---}}}}

\newcommand{\com}[1]       {}

\iflatexdiff

\renewcommand{\xout}[1]    {\textcolor{red}{\sout{#1}}}
 
\else

\newcommand{\xout}[1]    {}
\fi

\ifdraft
\usepackage{lineno}
\linenumbers
\modulolinenumbers[5]
\usepackage{fancyhdr}
\else
\renewcommand{\warn}[1]{}

\fi
\begin{document}
\newlength{\figlen}
\setlength{\figlen}{\linewidth}
\ifpreprint
\setlength{\figlen}{0.75\textwidth}
\begin{titlepage}
\PHyear{2016}
\PHnumber{229}                   
\PHdate{19 Sep}                  
\title{\papertitle}
\ShortTitle{\papertitle}
\Collaboration{ALICE Collaboration%
         \thanks{See Appendix~\ref{app:collab} for the list of collaboration members}}
\ShortAuthor{ALICE Collaboration} 
\ifdraft
\begin{center}
\today\\ \color{red}DRAFT \dvers\ \hspace{0.3cm} \$Revision: 3209 $\color{white}:$\$\color{black}\vspace{0.3cm}
\end{center}
\fi
\else
\title{\papertitle}
\iffull
\input{authors-prl.tex}
\else
\collaboration{ALICE Collaboration}
\fi
\vspace{0.3cm}
\ifdraft
\date{\today, \color{red}DRAFT \dvers\ \$Revision: 3209 $\color{white}:$\$\color{black}}
\else
\date{\today}
\fi
\fi
\begin{abstract}
\paperabstract
\end{abstract}
\ifpreprint
\end{titlepage}
\setcounter{page}{2}
\else
\maketitle
\fi
\ifdraft
\thispagestyle{fancyplain}
\fi

In elementary interactions with large momentum transfer ($Q^2 \gg \Lambda^2_{\rm QCD}$), partons with high transverse momentum ($\pt$) are produced. 
They evolve from high to low virtuality producing parton showers and eventually hadronizing into a spray of collimated
hadrons called jets. In interactions between heavy-ions, such high-\pt partons are produced at the early
stages of the collisions. They propagate through the dense and hot medium  created in these
collisions and are expected to lose energy due to medium-induced gluon radiation and elastic scatterings,
a process commonly referred to as jet quenching. Correspondingly, an inclusive jet suppression has been observed at the LHC \cite{Aad:2012vca, Abelev:2013kqa, Adam:2015ewa} together with a large di-jet energy asymmetry \cite{Aad:2010bu, Chatrchyan:2011sx}, while studies of the momentum and angular distributions of jet fragments show only a small modification of the jet core \cite{Chatrchyan:2012gw, Chatrchyan:2014ava, Aad:2014wha}, and an excess of soft particles radiated to large angles from the jet axis~\cite{Khachatryan:2016erx}.
Semi-inclusive hadron--jet correlations show a suppression of recoil jet yield, with no in-medium modification of transverse jet structure observed~\cite{Adam:2015doa}.

Di-hadron angular correlations represent a powerful complementary tool to study jet modifications on a statistical basis in an energy region where jets cannot be identified event-by-event over the fluctuating background.
Such studies involve measuring the distributions of the relative azimuthal angle $\Delta \varphi$
and pseudorapidity $\Delta \eta$ between particle pairs consisting of a trigger particle in a certain transverse momentum
$p_{\rm T ,trig}$ interval and an associated particle in a $p_{\rm T,assoc}$ interval.
In these correlations, jet production manifests itself as a peak centred around $(\Delta \varphi = 0,~\Delta \eta = 0 )$ (near-side peak) and a structure elongated in $\Delta\eta$ at $\Delta\varphi = \pi$ (the away-side or recoil-region). At low $\pt$, resonance decays as well as femtoscopic correlations also contribute to the near-side peak. The advantage of using di-hadron correlations is that an event-averaged subtraction of the background from particles uncorrelated to the jet can be performed.
This advantage is shared with the analysis of hadron-jet correlations recently reported in Ref.~\cite{Adam:2015doa,Khachatryan:2016erx}.

At RHIC, the near-side particle yield and peak shape of di-hadron correlations have been studied for different systems and collision
energies \cite{Abelev:2009af, Agakishiev:2011st, Adamczyk:2013jei}. Small modifications of the yields with respect to
a pp reference from PYTHIA are observed and there is remarkably little dependence on the collision system 
at the centre-of-mass energies of $\sqrt{s_{\rm NN}} = 62.4$ and $200$~GeV.
An exception is the measurement in central Au--Au collisions at $\sqrt{s_{\rm NN}} = 200$~GeV where the jet-like correlation is substantially broader
and the momentum spectrum softer than in peripheral collisions and than those in collisions of other systems in this kinematic
 regime. In Ref.~\cite{Agakishiev:2011st}, the broadening observed in central Au--Au collisions at \mbox{$\sqrt{s_{\rm NN}} = 200$~GeV} is seen as an indication of a modified jet fragmentation function. At the LHC, the measurement of the yield of particles associated to a high-\pt trigger 
 particle (\unit[8--15]{GeV/$c$}) in central Pb--Pb collisions relative to the pp reference at $p_{\rm T, assoc}>3$~GeV/$c$ shows a suppression
 on the away-side and a moderate enhancement on the near-side indicating that medium-induced  modifications 
 can also be expected on the near side \cite{Aamodt:2011vg}. Much stronger modification is observed for
 lower trigger and associated particle \pt\
($3 < p_{\rm T, trig} < 3.5$~GeV/$c$ and $1 < p_{\rm T, assoc} < 1.5$~GeV/$c$) \cite{Chatrchyan:2012wg,Adam:2016xbp}. In the most central Pb--Pb 
collisions, the near-side yield is enhanced by a factor of 1.7. The present paper expands 
these studies at the LHC to the characterisation of the angular distribution of the
associated particles with respect to the trigger particle. The angular distribution is sensitive to the broadening of
the jet due to its energy loss and the distribution of radiated energy.
Moreover, possible interactions of the parton shower with the collective longitudinal expansion~\cite{Armesto:2004pt, Armesto:2004vz, Romatschke:2006bb} or with turbulent colour fields~\cite{Majumder:2006wi} in the medium would result
in  near-side peak shapes that are broader in the $\Delta \eta$ than in the $\Delta \varphi$  direction.
Results from the study of the near-side peak shape of charged particles as a function of centrality and for different combinations of
trigger and associated particle \pt are discussed.

The data presented in this paper were taken by the ALICE detector, of which a detailed description can be found in Ref.~\cite{Aamodt:2008zz}.
The main subsystems used in the present analysis are the Inner Tracking System~(\ITS), and the
Time Projection Chamber~(\TPC). These have a common acceptance of
$|\eta| < 0.9$.
The \ITS\ consists of six layers of silicon detectors for vertex finding, tracking and triggering. 
The \TPC\ is the main tracking detector measuring up to 159 space points per track.
The V0 detector, two arrays of 32 scintillator tiles each, covering $2.8<\eta<5.1$~(V0-A) and $-3.7<\eta<-1.7$~(V0-C), was used for triggering and
centrality determination. All these detector systems have full azimuthal coverage.

Data from the 2010 and 2011 Pb--Pb runs of the LHC at $\snn=$~\unit[2.76]{TeV} are combined in the present analysis and it is compared with the 2011 pp run at the same energy. In total, about 39 million Pb--Pb and 30 million pp events are used. Details about the trigger and event selection in Pb--Pb (pp) collisions can be found in Ref.~\cite{Aamodt:2010cz} (Ref.~\cite{Abelev:2012sea}), while the centrality determination is described in Ref.~\cite{Abelev:2013qoq}. 

The collision-vertex position is determined with tracks reconstructed in the \ITS\ and \TPC~\cite{Abelev:2012hxa}, and its value in the beam direction ($z_{\rm vtx}$) is required to be within \unit[7]{cm} of the detector centre.
The Pb--Pb analysis is performed in the centrality classes 0--10\% (most central), 10--20\%, 20--30\%, 30--50\% and 50--80\%.
The analysis uses tracks reconstructed in the \ITS\ and \TPC\ with $1<\pt<$~\unit[8]{\gevc} and in a fiducial region of $|\eta|<0.8$. The track selection is described in Refs.~\cite{Abelev:2012ej, ALICE:2011ac}.
The efficiency and purity of the primary charged-particle selection are estimated from a Monte Carlo~(MC)
simulation using the HIJING~1.383 event generator~\cite{hijing} (for Pb--Pb) and the PYTHIA~6.4 event
generator~\cite{Sjostrand:2006za} with the tune Perugia-0~\cite{Skands:2010ak} (for \pp) with particle transport
through the detector using GEANT3~\cite{geant3ref2}.
The combined efficiency and acceptance for the track reconstruction in $|\eta|<0.8$ is
about 82--85\% at $\pt=$~\unit[1]{\gevc}, and decreases to about 76--80\% at $\pt=$~\unit[8]{\gevc} depending on collision system, data sample and event centrality.
The contamination originating from secondary particles from weak decays and  interactions in the detector material decreases from 2.5--4.5\% to 0.5--1\% in the $\pt$ range from 1 to \unit[8]{\gevc}.
The contribution from fake tracks is negligible. 
From these quantities a correction factor is computed as a function of $\eta$, $\pt$, $z_{\rm vtx}$ and event centrality which is applied as a weight for each trigger particle and particle pair in the analysis. 

The correlation between two charged particles (denoted trigger and associated particle) is measured
as a function of $\Dphi$ (defined within $-\pi/2$ and $3\pi/2$) and $\Deta$~\cite{alice_pa_ridge}.
The correlation is expressed in terms of the associated yield per trigger particle for intervals of $\ptt$ and $\pta$, measured as:  
\begin{equation}
\frac{1}{\Ntrig} \dNassoc = \frac{S(\Deta,\Dphi)}{B(\Deta,\Dphi)} \label{pertriggeryield}
\end{equation}
where $\Ntrig$ is the total number of trigger particles in the centrality class and $\ptt$ interval, ranging from 0.18 to 36 per event.
The signal distribution
$S(\Deta,\Dphi) = 1/\Ntrig\ \dd^2N_{\rm same}/\dd\Deta\dd\Dphi$
is the associated yield per trigger particle for particle pairs from the same event.
The background distribution \mbox{$B(\Deta,\Dphi) = \alpha\ \dd^2N_{\rm mixed}/\dd\Deta\dd\Dphi$}
accounts for the acceptance and efficiency of pair reconstruction.
It is constructed by correlating the trigger particles in one event with the associated particles from
other events.
The background distribution is scaled by a factor $\alpha$ which is chosen such that $B(0, 0)$ is unity for pairs where both particles travel in approximately the same direction (i.e.\ $\Dphi\approx 0,\ \Deta\approx 0$), and thus the efficiency and acceptance for the two particles are identical by construction.

A selection on the opening angle of the particle pairs is applied to both signal and background to avoid a bias due to the reduced efficiency for pairs with small opening angles. 
Furthermore, correlations induced by secondary particles from long-lived neutral-particle decays ($\rm K^0_s$ and $\Lambda$) and $\gamma$-conversions are suppressed by rejecting pairs in the corresponding invariant mass region.
A correction is performed for a mild $\Deta$ dependence of the structures in the two-particle correlation, which is due to a minor dependence of particle production and anisotropic flow on pseudorapidity. For further details on the analysis procedure, see the companion paper~\cite{Adam:2016ckp}.

In order to characterize the near-side peak shape, a simultaneous fit of the peak, the combinatorial background and the long-range correlation background stemming from collective effects is performed. This exploits that in two-particle correlations the near-side peak is centred around $\Dphi = 0,~\Deta = 0$, while long-range correlation structures are mostly independent of $\Deta$~\cite{Aad:2014eoa}. The away-side peak, however, is elongated in $\Deta$, therefore this strategy cannot be applied to studying the away side. The fit function used is a combination of a constant, a generalized two-dimensional Gaussian function and $\cos( n \Dphi)$ terms for $n = 2, 3, 4$.
\begin{eqnarray}
  F(\Dphi,\Deta) &=& C_1 + \sum_{n=2}^4 2 V_{n\Delta} \cos (n \Dphi) + \nonumber \\
    && C_2 \cdot G_{\gamma_{\Dphi},w_{\Dphi}}(\Dphi) \cdot G_{\gamma_{\Deta},w_{\Deta}}(\Deta) \label{eq:fit} \\
    G_{\gamma_x,w_x}(x) &=& \frac{\gamma_x}{2w_x\Gamma (1/\gamma_x)} \exp \left[ -\left(\frac{|x|}{w_x}\right)^{\gamma_x} \right].
\end{eqnarray}
Thus in Pb--Pb collisions, the background is characterized by four parameters ($C_1$, $V_{n\Delta}$) where $V_{n\Delta}$ are the Fourier components of the long-range correlations~\cite{Aamodt:2011by}, and it should be noted that the inclusion of  orders higher than four does not significantly change the fit results. In pp collisions, the background consists effectively only of the pedestal $C_1$. The peak magnitude is characterized by $C_2$, and the shape, which is the focus of the present analysis, by four parameters ($\gamma_{\Dphi},~w_{\Dphi},~\gamma_{\Deta},~w_{\Deta}$). The aim of using this fit function is to allow for a compact description of the data rather than attempting to give a physical meaning to each parameter. Therefore, the variance of $G$ is calculated, which reduces the description of the peak shape to two parameters ($\sigma_{\Dphi}$ and $\sigma_{\Deta}$). To describe the evolution of the peak shape from peripheral to central collisions the ratios of the widths in the central bin (0--10\%) and the peripheral bin (50--80\%), denoted by $\sigma^{\rm CP}_{\Dphi}$ and $\sigma^{\rm CP}_{\Deta}$, are also calculated.

In the data a depletion around $\Dphi = 0$, $\Deta = 0$ is observed at low $\pt$, however, the fit function does not include such a depletion. To avoid a bias on the extracted peak width, some bins in the central region are excluded from the fit. The size of the excluded region varies with $\pt$ and collision centrality (from no exclusion to 0.3). Thus, by definition, the peak width describes the shape of the peak outside of the central region. The depletion in the central region is quantified below by computing the difference between the fit and the per-trigger yield within the exclusion region.

In Pb--Pb collisions, the obtained $\chi^2/{\rm ndf}$ values of the fits are in the range 1.0--2.5; most are around 1.5. In the highest two $\pt$ bins (i.e. in $3<\pta<$~\unit[8]{\gevc} and $4<\ptt<$~\unit[8]{\gevc}) the values increase up to about 2.5 showing that at high $\pt$ the peak shape starts to depart from the generalized Gaussian description. In pp collisions, the $\chi^2/{\rm ndf}$ values are in the range 1.3--2.0.

\begin{figure}[t!]
\centering
\includegraphics[width=0.47\figlen]{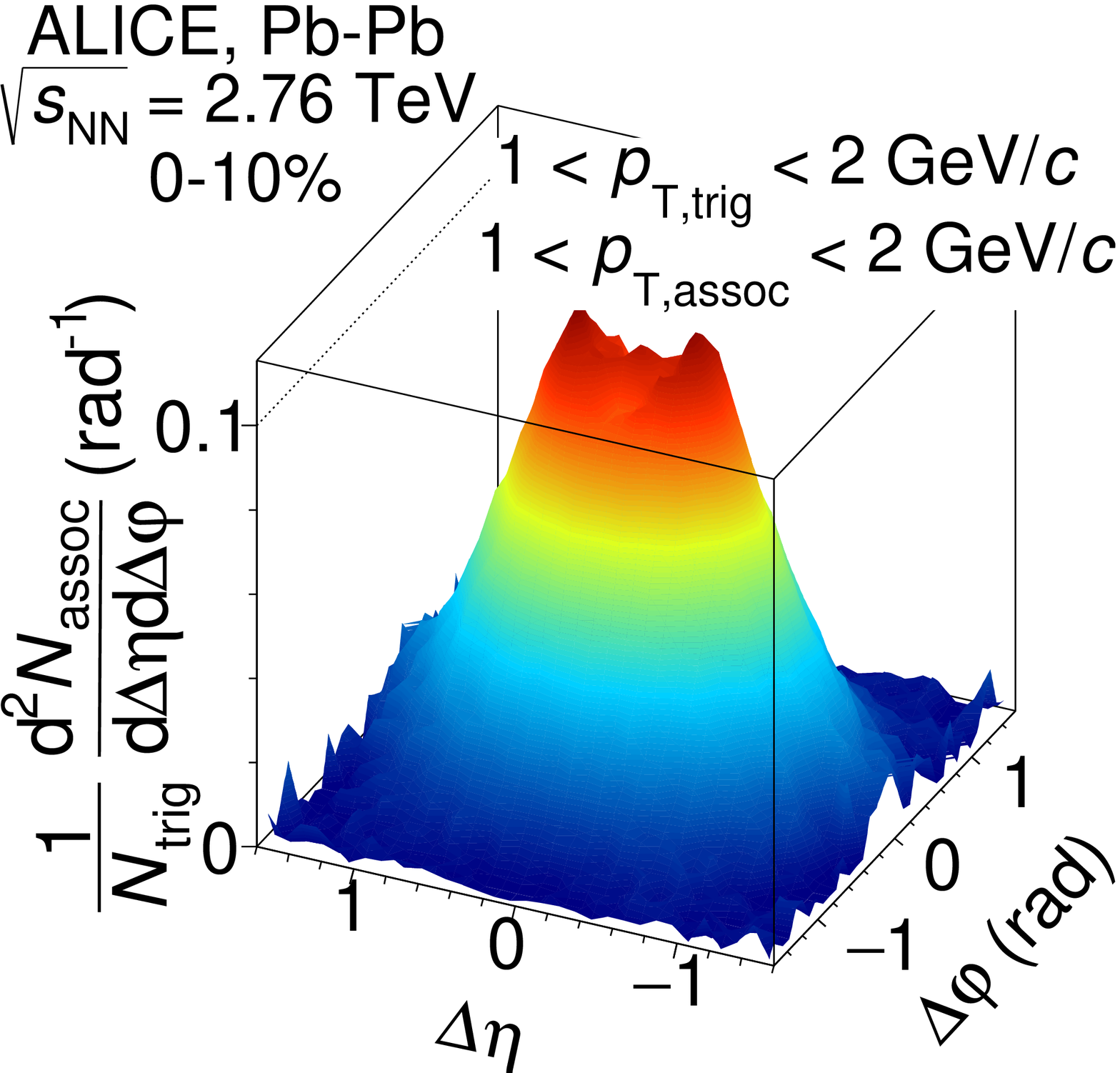}
\includegraphics[width=0.51\figlen]{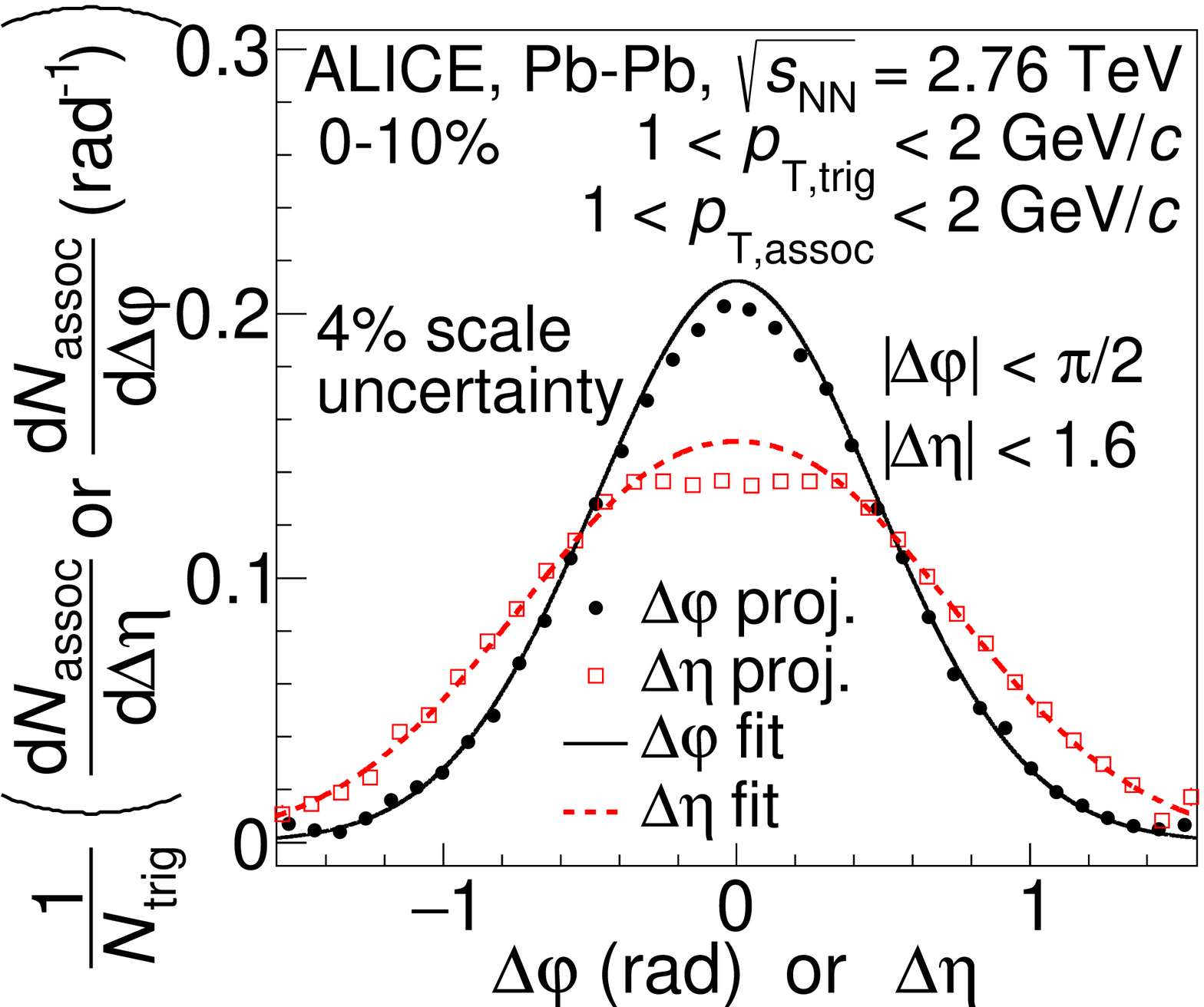}
\caption{\label{fig:pertriggeryields}
Left panel: associated yield per trigger particle as a function of $\Dphi$ and $\Deta$. The background obtained from the fit function has been subtracted in order to emphasize the near-side peak. Right panel: projections to the $\Dphi$ and $\Deta$ axes overlaid with the peak part of the fit function.}
\end{figure}

Systematic uncertainties connected to the measurement are determined by modifying
the event and track selections. In addition, uncertainties related to the cut on pairs with small opening angles and neutral-particle decays, as well as the sensitivity to the pseudorapidity range are considered. The difference in the extracted parameters is studied as a function of $\pt$, centrality and collision system, but these dependencies are rather weak and in most cases one uncertainty value can be quoted for each type of systematic uncertainty. Finally, the different sources of systematic uncertainties are added in quadrature. The extracted peak widths are rather insensitive to changes in the selections (total uncertainty of about 2--4.5\%), while the near-side depletion yield is more sensitive (about 24--45\% uncertainty).
The contribution from resonance decays was studied by performing the analysis separately for like and unlike sign pairs, and a significant influence on the results presented below was not found.

Figure~\ref{fig:pertriggeryields} shows the near-side peak in $1<\ptt<$~\unit[2]{\gevc} and $1<\pta<$~\unit[2]{\gevc} for the 10\% most central collisions. In addition to the two-dimensional representation, projections are shown where the background estimated with Eq.~\ref{eq:fit} has been subtracted. The near-side peak is asymmetric, i.e. wider in $\Deta$ than in $\Dphi$. It is also broader than in peripheral Pb--Pb and pp collisions where it is mostly symmetric in $\Dphi$ and $\Deta$ (not shown, see Ref.~\cite{Adam:2016ckp}).
Furthermore, a depletion around $\Dphi = 0$, $\Deta = 0$ develops which will be discussed in more detail further below.
Also at higher $\pt$, the near-side peak is broader in central collisions than in peripheral or pp collisions. This broadening is less pronounced at high $\pt$ than at low $\pt$, and the asymmetry between $\Dphi$ and $\Deta$ disappears at the two highest $\pt$ bins, 
see Ref.~\cite{Adam:2016ckp}.

\begin{figure}[t!]
\centering
\includegraphics[width=\figlen,trim=0 0 35 0,clip=true]{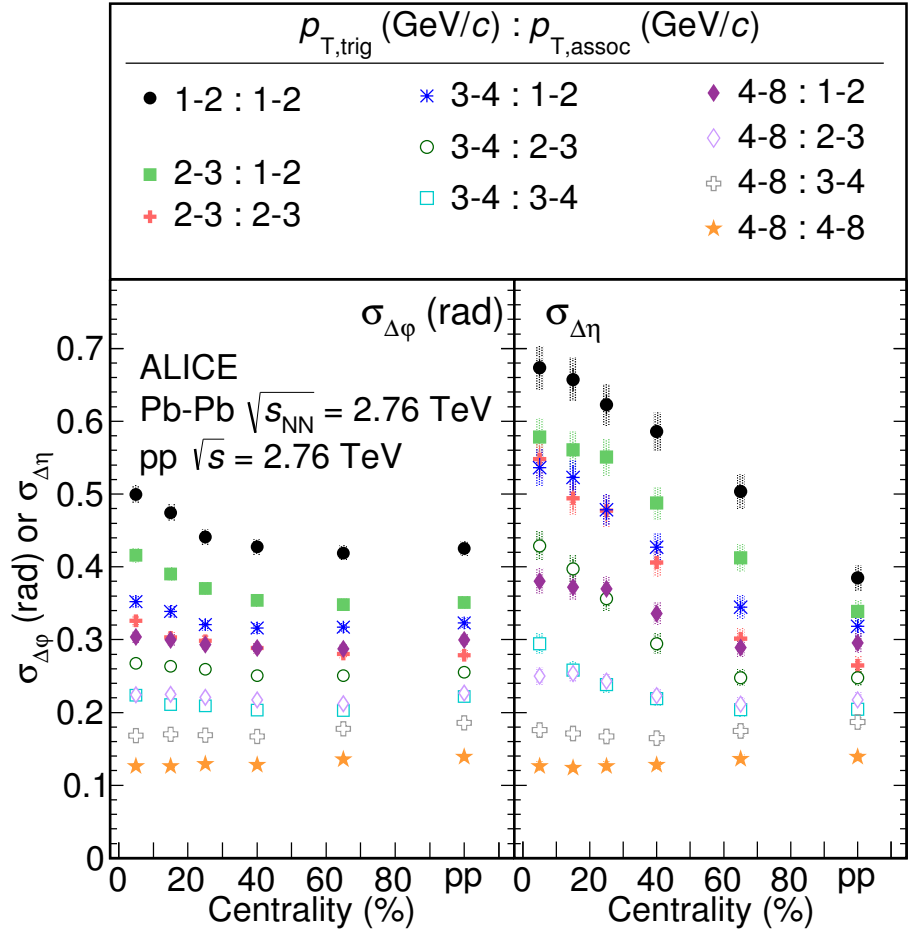}
\caption{\label{fig:sigma}
Shape parameters $\sigma_{\Dphi}$ (left panel) and $\sigma_{\Deta}$ (right panel) as a function of centrality in different $\pt$ ranges. Lines indicate statistical uncertainties (mostly smaller than the marker size), while boxes denote systematic uncertainties. The markers are placed at the centre of the centrality bins.
}
\end{figure}
\begin{figure}[t!]
\centering
\includegraphics[width=\figlen]{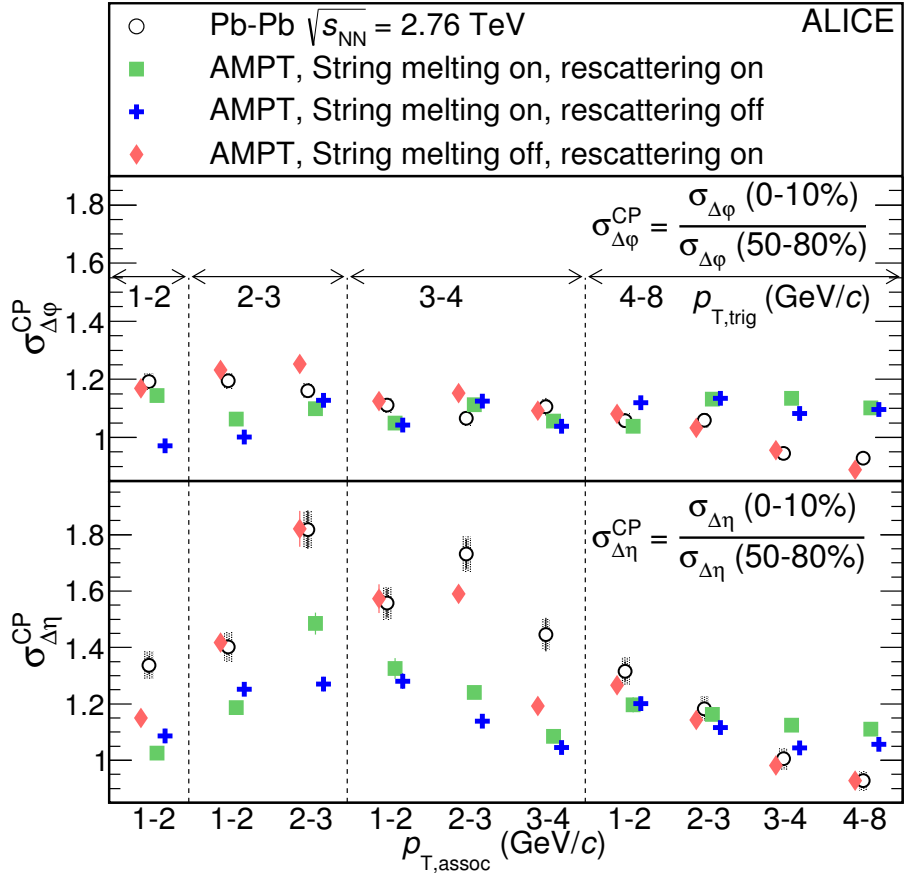}
\caption{\label{fig:widthratios}
Ratio of the peak widths in $\Dphi$ (top) and $\Deta$ (bottom) observed in central (0-10\%) and peripheral (50-80\%) collisions as a function of $\pt _{\rm ,trig}$ and $\pt _{\rm ,assoc}$ ranges. The data is compared to different AMPT settings. Note that the $x$-axis combines the $\pta$ and $\ptt$ axis, and therefore, a uniform trend of the values is not expected. 
}
\end{figure}

The extracted shape parameters $\sigma_{\Dphi}$ and $\sigma_{\Deta}$ are presented in Fig.~\ref{fig:sigma}. In pp collisions, the $\sigma$ values range from about 0.14 to about 0.43 showing a $\pt$ dependence qualitatively expected due to the boost of the evolving parton shower: at larger $\pt$ the peak is narrower.
In the $\Dphi$ direction (left panel) the values obtained in pp collisions are consistent with those in peripheral Pb--Pb collisions. The peak width increases towards central events, which is most pronounced in the lowest $\pt$ bin (20\% increase). In the higher $\pt$ bins no significant width increase can be observed.
In the $\Deta$ direction (right panel) a much larger broadening is found towards central collisions. Already in peripheral collisions the width is larger than in pp collisions. From peripheral to central collisions the width increases further up to $\sigma_{\Deta} = 0.67$ in the lowest $\pt$ bin and the largest relative increase of about 85\% is observed for $2<\ptt<$~\unit[3]{\gevc} and $2<\pta<$~\unit[3]{\gevc}. For all but the two largest $\pt$ bins a significant broadening can be observed. This increase is quantified for all $\pt$ bins in Fig.~\ref{fig:widthratios} by $\sigma^{\rm CP}_{\Dphi}$ and $\sigma^{\rm CP}_{\Deta}$. The increase is quantified with respect to peripheral Pb--Pb instead of pp collisions to facilitate the MC comparisons discussed below.

In pp collisions, the peak shows circular symmetry in the $\Deta$--$\Dphi$ plane for all $\pt$ values. In Pb--Pb collisions, the peak becomes asymmetric towards central collisions for all but the two highest $\pt$ bins. The magnitude of this asymmetry depends on $\pt$ and is largest with about 70\% ($\sigma_{\Deta} > \sigma_{\Dphi}$) in the range $2<\pta<$~\unit[3]{\gevc} and $2<\ptt<$~\unit[3]{\gevc}.
These results are compatible with a similar study by the STAR collaboration at $\sqrt{s_{\rm NN}} = \unit[200]{GeV}$~\cite{Agakishiev:2011st}, which is detailed in the companion paper~\cite{Adam:2016ckp}.

In Ref.~\cite{Armesto:2004pt} it was suggested that the interplay of longitudinal flow with a fragmenting high $\pt$ parton can lead to the observed asymmetric peak shape.
The authors argue that hard partons interact with a medium which shows collective behaviour, contrary to the simpler picture where the parton propagates through an isotropic medium with respect to the parton direction. In their calculation the scattering centres are Lorentz-boosted by applying a momentum shift depending on the collective component transverse to the parton-propagation direction.
The calculation in Ref.~\cite{Armesto:2004pt} for Au--Au collisions at \mbox{\snn\ = \unit[200]{GeV}} predicts a 20\% increase from peripheral to central events in the $\Dphi$ direction and a 60\% increase in the $\Deta$ direction, which is in good agreement with the measurements by the STAR collaboration. Despite the different centre of mass energy and collision system, the calculation is in quantitative agreement with the results presented in this paper as well.

In order to study further the possibility that an interplay of flow and jets can cause the observation, the data is compared to results from A Multi-Phase Transport model (AMPT)~\cite{Lin:2004en, Xu:2011fi}. Two mechanisms in AMPT produce collective effects: partonic and hadronic rescattering. 
Before partonic rescattering, the initially produced strings may be broken into smaller pieces by the so-called string melting.
Three different AMPT settings are considered which have either string melting (configuration (a)) or hadronic rescattering (b) or both activated (c)~\cite{Adam:2016nfo}. 

The peak widths are extracted from particle-level AMPT simulations in the same way as for the data. None of the AMPT settings provides an accurate description of the measured absolute widths. Further discussion and the corresponding figure can be found in Ref.~\cite{Adam:2016ckp}. In order to provide nevertheless a meaningful comparison of the relative increase, $\sigma^{\rm CP}_{\Dphi}$ and $\sigma^{\rm CP}_{\Deta}$ from the models are shown together with the data in Fig.~\ref{fig:widthratios}.
In the $\Dphi$ direction, the setting with string melting deactivated and hadronic rescattering active follows the trend of the data closest. The two other settings show a more uniform distribution across $\pt$ and only differ in the two lowest $\pt$ bins.
In the $\Deta$ direction, the setting with string melting deactivated and hadronic rescattering active quite remarkably follows the trend of the data including the large increase for intermediate $\pt$. The two other settings show qualitatively a similar trend but miss the data quantitatively.

The presented results have focused up to now on the overall shape of the near-side peak. In addition to the broadening, a distinct feature is observed, a depletion around $\Dphi = 0$, $\Deta = 0$ (see Fig.~\ref{fig:pertriggeryields}).
An extensive set of studies was carried out to exclude that this depletion could arise from a detector effect. A similar structure is found in AMPT simulations with hadronic rescattering regardless of the string melting setting~\cite{Adam:2016ckp}.

\begin{figure}[t!]
\centering
\includegraphics[width=\figlen]{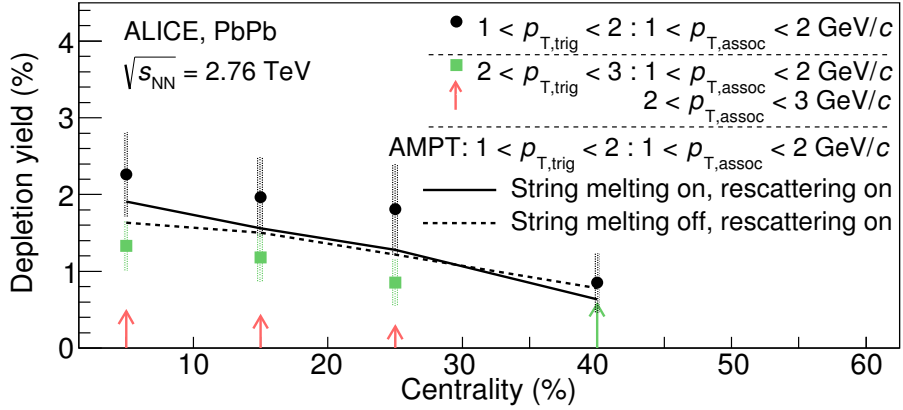}
\caption{\label{fig:depletionquantification}
Missing yield in the depletion region relative to the overall peak yield extracted from the fit. The arrows indicate the upper limit in case the uncertainty bands touch 0. For comparison, the non-zero values from two AMPT simulations are shown as lines.}
\end{figure}

In order to quantify this depletion, the difference is computed between the fit (where the depletion region has been excluded, see above) and the per-trigger yield for each $\pt$ bin. This is normalized by the total peak yield and it is referred to as depletion yield in the following. The region, where effects are expected from the limited two-track reconstruction efficiency ($|\Dphi| < 0.04$ and $|\Deta| < 0.05$, which corresponds to 0.5--6\% of the integrated region), is excluded from this calculation. Fig.~\ref{fig:depletionquantification} presents the depletion yield as a function of centrality for the $\pt$ bins, where it is different from 0. It can be seen that (2.2$\pm$0.5)\% of the yield is missing in the lowest $\pt$ bin and in the 10\% most central events. This value decreases gradually with centrality and with $\pt$. No significant depletion yield is observed for 50--80\% (30--80\%) centrality or pp collisions for the lowest (second lowest) $\pt$ range. 
The depletion observed in the AMPT events is present only in the lowest $\pt$ bin, where its value is compatible with the data for both settings where hadronic rescattering is switched on. For larger $\pt$ bins and for the configuration without hadronic rescattering the depletion yield is consistent with 0 in AMPT.

The reported results can be interpreted in the context of radial and anisotropic flow by calculating the radial-flow expansion velocity $\beta_{\rm T}$ and the elliptic flow coefficient $v_2\{2\}$ from the 10\% most central events from data and from the AMPT samples. The expansion velocity $\beta_{\rm T}$ is extracted from a Blast-Wave fit to the $\pt$-spectra of $\pi$, K and p in the rapidity range of $|y| < 0.5$~\cite{Abelev:2013vea}. The $v_2\{2\}$ is extracted from two-particle correlations within $|\eta| < 0.8$ and $0.2 < \pt <$~\unit[5]{\gevc}~\cite{Aamodt:2010pa}.

The depletion (Fig.~\ref{fig:depletionquantification}) occurs in the two AMPT configurations (b) and (c) where the $\beta_{\rm T}$ is large, while the configuration (a) without the depletion has the smallest $\beta_{\rm T}$.
The coefficient $v_2\{2\}$ has significantly different values in the two configurations (b) and (c) with depletion , and the relative increase of the peak width (Fig.~\ref{fig:widthratios}) is best described by the AMPT configuration with the largest $\beta_{\rm T}$ (b).
Based on these studies, it seems that the depletion and the broadening observed in the data are more likely accompanied by radial flow than elliptic flow.

In summary, we have presented a detailed characterization of the flow-subtracted near-side peak in two-particle correlations in Pb--Pb collisions at $\snn=$~\unit[2.76]{TeV} together with a measurement in pp collisions at the same energy. The near-side peak shows several untypical characteristics in Pb--Pb collisions: the peak gets broader and more asymmetric from peripheral towards central collisions over a wide $\pt$ range, and an unexpected depletion develops in central collisions at low $\pt$. The broadening is present both in the $\Dphi$ and the $\Deta$ directions, but it is significantly stronger in the $\Deta$ direction, leading to the asymmetric shape of the peak. The near-side peak also shows a characteristic $\pt$ dependence in both Pb--Pb and pp collisions.

AMPT simulations show also an asymmetric broadening, and the depletion is present when hadronic rescattering is included. The AMPT configuration with hadronic rescattering and without string melting reproduces quantitatively the relative peak broadening as well as the size of the depletion, underlining the importance of the hadronic phase in heavy-ion collisions.
The extraction of the radial-flow expansion velocity suggests that the stronger the radial flow, the stronger the observed effects are. In addition, earlier theoretical and phenomenological work connected the longitudinal broadening of the near-side jet-like peak to strong longitudinal flow in AMPT~\cite{Ma:2008nd}, as well as to an interplay of partons traversing the longitudinally expanding medium~\cite{Armesto:2004pt}. Thus a possible scenario is that
the presented observations are caused by the interplay of the jet with the collective expansion.

\ifpreprint
\iffull
\newenvironment{acknowledgement}{\relax}{\relax}
\begin{acknowledgement}
\section*{Acknowledgements}

The ALICE Collaboration would like to thank all its engineers and technicians for their invaluable contributions to the construction of the experiment and the CERN accelerator teams for the outstanding performance of the LHC complex.
The ALICE Collaboration gratefully acknowledges the resources and support provided by all Grid centres and the Worldwide LHC Computing Grid (WLCG) collaboration.
The ALICE Collaboration acknowledges the following funding agencies for their support in building and running the ALICE detector:
A. I. Alikhanyan National Science Laboratory (Yerevan Physics Institute) Foundation (ANSL), State Committee of Science and World Federation of Scientists (WFS), Armenia;
Austrian Academy of Sciences and Nationalstiftung f\"{u}r Forschung, Technologie und Entwicklung, Austria;
, Conselho Nacional de Desenvolvimento Cient\'{\i}fico e Tecnol\'{o}gico (CNPq), Financiadora de Estudos e Projetos (Finep) and Funda\c{c}\~{a}o de Amparo \`{a} Pesquisa do Estado de S\~{a}o Paulo (FAPESP), Brazil;
Ministry of Education of China (MOE of China), Ministry of Science \& Technology of China (MOST of China) and National Natural Science Foundation of China (NSFC), China;
Ministry of Science, Education and Sport and Croatian Science Foundation, Croatia;
Centro de Investigaciones Energ\'{e}ticas, Medioambientales y Tecnol\'{o}gicas (CIEMAT), Cuba;
Ministry of Education, Youth and Sports of the Czech Republic, Czech Republic;
Danish National Research Foundation (DNRF), The Carlsberg Foundation and The Danish Council for Independent Research | Natural Sciences, Denmark;
Helsinki Institute of Physics (HIP), Finland;
Commissariat \`{a} l'Energie Atomique (CEA) and Institut National de Physique Nucl\'{e}aire et de Physique des Particules (IN2P3) and Centre National de la Recherche Scientifique (CNRS), France;
Bundesministerium f\"{u}r Bildung, Wissenschaft, Forschung und Technologie (BMBF) and GSI Helmholtzzentrum f\"{u}r Schwerionenforschung GmbH, Germany;
Ministry of Education, Research and Religious Affairs, Greece;
National Research, Development and Innovation Office, Hungary;
Department of Atomic Energy Government of India (DAE), India;
Indonesian Institute of Science, Indonesia;
Centro Fermi - Museo Storico della Fisica e Centro Studi e Ricerche Enrico Fermi and Istituto Nazionale di Fisica Nucleare (INFN), Italy;
Institute for Innovative Science and Technology , Nagasaki Institute of Applied Science (IIST), Japan Society for the Promotion of Science (JSPS) KAKENHI and Japanese Ministry of Education, Culture, Sports, Science and Technology (MEXT), Japan;
Consejo Nacional de Ciencia (CONACYT) y Tecnolog\'{i}a, through Fondo de Cooperaci\'{o}n Internacional en Ciencia y Tecnolog\'{i}a (FONCICYT) and Direcci\'{o}n General de Asuntos del Personal Academico (DGAPA), Mexico;
Nationaal instituut voor subatomaire fysica (Nikhef), Netherlands;
The Research Council of Norway, Norway;
Commission on Science and Technology for Sustainable Development in the South (COMSATS), Pakistan;
Pontificia Universidad Cat\'{o}lica del Per\'{u}, Peru;
Ministry of Science and Higher Education and National Science Centre, Poland;
Ministry of Education and Scientific Research, Institute of Atomic Physics and Romanian National Agency for Science, Technology and Innovation, Romania;
Joint Institute for Nuclear Research (JINR), Ministry of Education and Science of the Russian Federation and National Research Centre Kurchatov Institute, Russia;
Ministry of Education, Science, Research and Sport of the Slovak Republic, Slovakia;
National Research Foundation of South Africa, South Africa;
Korea Institute of Science and Technology Information and National Research Foundation of Korea (NRF), South Korea;
Centro de Investigaciones Energ\'{e}ticas, Medioambientales y Tecnol\'{o}gicas (CIEMAT) and Ministerio de Ciencia e Innovacion, Spain;
Knut \& Alice Wallenberg Foundation (KAW) and Swedish Research Council (VR), Sweden;
European Organization for Nuclear Research, Switzerland;
National Science and Technology Development Agency (NSDTA), Office of the Higher Education Commission under NRU project of Thailand and Suranaree University of Technology (SUT), Thailand;
Turkish Atomic Energy Agency (TAEK), Turkey;
National Academy of  Sciences of Ukraine, Ukraine;
Science and Technology Facilities Council (STFC), United Kingdom;
National Science Foundation of the United States of America (NSF) and United States Department of Energy, Office of Nuclear Physics (DOE NP), United States.    
\end{acknowledgement}
\fi
\ifbibtex
\bibliographystyle{utphys}
\bibliography{biblio}{}
\else
\providecommand{\href}[2]{#2}\begingroup\raggedright\endgroup

\fi
\iffull
\newpage
\appendix
\section{The ALICE Collaboration}
\label{app:collab}



\begingroup
\small
\begin{flushleft}
J.~Adam$^\textrm{\scriptsize 39}$,
D.~Adamov\'{a}$^\textrm{\scriptsize 86}$,
M.M.~Aggarwal$^\textrm{\scriptsize 90}$,
G.~Aglieri Rinella$^\textrm{\scriptsize 35}$,
M.~Agnello$^\textrm{\scriptsize 113}$\textsuperscript{,}$^\textrm{\scriptsize 31}$,
N.~Agrawal$^\textrm{\scriptsize 48}$,
Z.~Ahammed$^\textrm{\scriptsize 137}$,
S.~Ahmad$^\textrm{\scriptsize 18}$,
S.U.~Ahn$^\textrm{\scriptsize 70}$,
S.~Aiola$^\textrm{\scriptsize 141}$
% ...
\renewcommand\labelenumi{\textsuperscript{\theenumi}~}

\section*{Affiliation notes}
\renewcommand\theenumi{\roman{enumi}}
\begin{Authlist}
\item \Adef{0}Deceased
\item \Adef{idp26411852}{Also at: Georgia State University, Atlanta, Georgia, United States}
\item \Adef{idp27127172}{Also at: Also at Department of Applied Physics, Aligarh Muslim University, Aligarh, India}
\item \Adef{idp27505124}{Also at: M.V. Lomonosov Moscow State University, D.V. Skobeltsyn Institute of Nuclear, Physics, Moscow, Russia}
\end{Authlist}

\section*{Collaboration Institutes}
\renewcommand\theenumi{\arabic{enumi}~}

$^{1}$A.I. Alikhanyan National Science Laboratory (Yerevan Physics Institute) Foundation, Yerevan, Armenia
\\
$^{2}$Benem\'{e}rita Universidad Aut\'{o}noma de Puebla, Puebla, Mexico
\\
$^{3}$Bogolyubov Institute for Theoretical Physics, Kiev, Ukraine
\\
$^{4}$Bose Institute, Department of Physics
and Centre for Astroparticle Physics and Space Science (CAPSS), Kolkata, India
\\
$^{5}$Budker Institute for Nuclear Physics, Novosibirsk, Russia
\\
$^{6}$California Polytechnic State University, San Luis Obispo, California, United States
\\
$^{7}$Central China Normal University, Wuhan, China
\\
$^{8}$Centre de Calcul de l'IN2P3, Villeurbanne, Lyon, France
\\
$^{9}$Centro de Aplicaciones Tecnol\'{o}gicas y Desarrollo Nuclear (CEADEN), Havana, Cuba
\\
$^{10}$Centro de Investigaciones Energ\'{e}ticas Medioambientales y Tecnol\'{o}gicas (CIEMAT), Madrid, Spain
\\
$^{11}$Centro de Investigaci\'{o}n y de Estudios Avanzados (CINVESTAV), Mexico City and M\'{e}rida, Mexico
\\
$^{12}$Centro Fermi - Museo Storico della Fisica e Centro Studi e Ricerche ``Enrico Fermi', Rome, Italy
\\
$^{13}$Chicago State University, Chicago, Illinois, United States
\\
$^{14}$China Institute of Atomic Energy, Beijing, China
\\
$^{15}$Commissariat \`{a} l'Energie Atomique, IRFU, Saclay, France
\\
$^{16}$COMSATS Institute of Information Technology (CIIT), Islamabad, Pakistan
\\
$^{17}$Departamento de F\'{\i}sica de Part\'{\i}culas and IGFAE, Universidad de Santiago de Compostela, Santiago de Compostela, Spain
\\
$^{18}$Department of Physics, Aligarh Muslim University, Aligarh, India
\\
$^{19}$Department of Physics, Ohio State University, Columbus, Ohio, United States
\\
$^{20}$Department of Physics, Sejong University, Seoul, South Korea
\\
$^{21}$Department of Physics, University of Oslo, Oslo, Norway
\\
$^{22}$Department of Physics and Technology, University of Bergen, Bergen, Norway
\\
$^{23}$Dipartimento di Fisica dell'Universit\`{a} 'La Sapienza'
and Sezione INFN, Rome, Italy
\\
$^{24}$Dipartimento di Fisica dell'Universit\`{a}
and Sezione INFN, Cagliari, Italy
\\
$^{25}$Dipartimento di Fisica dell'Universit\`{a}
and Sezione INFN, Trieste, Italy
\\
$^{26}$Dipartimento di Fisica dell'Universit\`{a}
and Sezione INFN, Turin, Italy
\\
$^{27}$Dipartimento di Fisica e Astronomia dell'Universit\`{a}
and Sezione INFN, Bologna, Italy
\\
$^{28}$Dipartimento di Fisica e Astronomia dell'Universit\`{a}
and Sezione INFN, Catania, Italy
\\
$^{29}$Dipartimento di Fisica e Astronomia dell'Universit\`{a}
and Sezione INFN, Padova, Italy
\\
$^{30}$Dipartimento di Fisica `E.R.~Caianiello' dell'Universit\`{a}
and Gruppo Collegato INFN, Salerno, Italy
\\
$^{31}$Dipartimento DISAT del Politecnico and Sezione INFN, Turin, Italy
\\
$^{32}$Dipartimento di Scienze e Innovazione Tecnologica dell'Universit\`{a} del Piemonte Orientale and INFN Sezione di Torino, Alessandria, Italy
\\
$^{33}$Dipartimento Interateneo di Fisica `M.~Merlin'
and Sezione INFN, Bari, Italy
\\
$^{34}$Division of Experimental High Energy Physics, University of Lund, Lund, Sweden
\\
$^{35}$European Organization for Nuclear Research (CERN), Geneva, Switzerland
\\
$^{36}$Excellence Cluster Universe, Technische Universit\"{a}t M\"{u}nchen, Munich, Germany
\\
$^{37}$Faculty of Engineering, Bergen University College, Bergen, Norway
\\
$^{38}$Faculty of Mathematics, Physics and Informatics, Comenius University, Bratislava, Slovakia
\\
$^{39}$Faculty of Nuclear Sciences and Physical Engineering, Czech Technical University in Prague, Prague, Czech Republic
\\
$^{40}$Faculty of Science, P.J.~\v{S}af\'{a}rik University, Ko\v{s}ice, Slovakia
\\
$^{41}$Faculty of Technology, Buskerud and Vestfold University College, Tonsberg, Norway
\\
$^{42}$Frankfurt Institute for Advanced Studies, Johann Wolfgang Goethe-Universit\"{a}t Frankfurt, Frankfurt, Germany
\\
$^{43}$Gangneung-Wonju National University, Gangneung, South Korea
\\
$^{44}$Gauhati University, Department of Physics, Guwahati, India
\\
$^{45}$Helmholtz-Institut f\"{u}r Strahlen- und Kernphysik, Rheinische Friedrich-Wilhelms-Universit\"{a}t Bonn, Bonn, Germany
\\
$^{46}$Helsinki Institute of Physics (HIP), Helsinki, Finland
\\
$^{47}$Hiroshima University, Hiroshima, Japan
\\
$^{48}$Indian Institute of Technology Bombay (IIT), Mumbai, India
\\
$^{49}$Indian Institute of Technology Indore, Indore, India
\\
$^{50}$Indonesian Institute of Sciences, Jakarta, Indonesia
\\
$^{51}$Inha University, Incheon, South Korea
\\
$^{52}$Institut de Physique Nucl\'eaire d'Orsay (IPNO), Universit\'e Paris-Sud, CNRS-IN2P3, Orsay, France
\\
$^{53}$Institute for Nuclear Research, Academy of Sciences, Moscow, Russia
\\
$^{54}$Institute for Subatomic Physics of Utrecht University, Utrecht, Netherlands
\\
$^{55}$Institute for Theoretical and Experimental Physics, Moscow, Russia
\\
$^{56}$Institute of Experimental Physics, Slovak Academy of Sciences, Ko\v{s}ice, Slovakia
\\
$^{57}$Institute of Physics, Academy of Sciences of the Czech Republic, Prague, Czech Republic
\\
$^{58}$Institute of Physics, Bhubaneswar, India
\\
$^{59}$Institute of Space Science (ISS), Bucharest, Romania
\\
$^{60}$Institut f\"{u}r Informatik, Johann Wolfgang Goethe-Universit\"{a}t Frankfurt, Frankfurt, Germany
\\
$^{61}$Institut f\"{u}r Kernphysik, Johann Wolfgang Goethe-Universit\"{a}t Frankfurt, Frankfurt, Germany
\\
$^{62}$Institut f\"{u}r Kernphysik, Westf\"{a}lische Wilhelms-Universit\"{a}t M\"{u}nster, M\"{u}nster, Germany
\\
$^{63}$Instituto de Ciencias Nucleares, Universidad Nacional Aut\'{o}noma de M\'{e}xico, Mexico City, Mexico
\\
$^{64}$Instituto de F\'{i}sica, Universidade Federal do Rio Grande do Sul (UFRGS), Porto Alegre, Brazil
\\
$^{65}$Instituto de F\'{\i}sica, Universidad Nacional Aut\'{o}noma de M\'{e}xico, Mexico City, Mexico
\\
$^{66}$Institut Pluridisciplinaire Hubert Curien (IPHC), Universit\'{e} de Strasbourg, CNRS-IN2P3, Strasbourg, France, Strasbourg, France
\\
$^{67}$iThemba LABS, National Research Foundation, Somerset West, South Africa
\\
$^{68}$Joint Institute for Nuclear Research (JINR), Dubna, Russia
\\
$^{69}$Konkuk University, Seoul, South Korea
\\
$^{70}$Korea Institute of Science and Technology Information, Daejeon, South Korea
\\
$^{71}$KTO Karatay University, Konya, Turkey
\\
$^{72}$Laboratoire de Physique Corpusculaire (LPC), Clermont Universit\'{e}, Universit\'{e} Blaise Pascal, CNRS--IN2P3, Clermont-Ferrand, France
\\
$^{73}$Laboratoire de Physique Subatomique et de Cosmologie, Universit\'{e} Grenoble-Alpes, CNRS-IN2P3, Grenoble, France
\\
$^{74}$Laboratori Nazionali di Frascati, INFN, Frascati, Italy
\\
$^{75}$Laboratori Nazionali di Legnaro, INFN, Legnaro, Italy
\\
$^{76}$Lawrence Berkeley National Laboratory, Berkeley, California, United States
\\
$^{77}$Moscow Engineering Physics Institute, Moscow, Russia
\\
$^{78}$Nagasaki Institute of Applied Science, Nagasaki, Japan
\\
$^{79}$National Centre for Nuclear Studies, Warsaw, Poland
\\
$^{80}$National Institute for Physics and Nuclear Engineering, Bucharest, Romania
\\
$^{81}$National Institute of Science Education and Research, Bhubaneswar, India
\\
$^{82}$National Research Centre Kurchatov Institute, Moscow, Russia
\\
$^{83}$Niels Bohr Institute, University of Copenhagen, Copenhagen, Denmark
\\
$^{84}$Nikhef, Nationaal instituut voor subatomaire fysica, Amsterdam, Netherlands
\\
$^{85}$Nuclear Physics Group, STFC Daresbury Laboratory, Daresbury, United Kingdom
\\
$^{86}$Nuclear Physics Institute, Academy of Sciences of the Czech Republic, \v{R}e\v{z} u Prahy, Czech Republic, \v{R}e\v{z} u Prahy, Czech Republic
\\
$^{87}$Oak Ridge National Laboratory, Oak Ridge, Tennessee, United States
\\
$^{88}$Petersburg Nuclear Physics Institute, Gatchina, Russia
\\
$^{89}$Physics Department, Creighton University, Omaha, Nebraska, United States
\\
$^{90}$Physics Department, Panjab University, Chandigarh, India
\\
$^{91}$Physics Department, University of Athens, Athens, Greece
\\
$^{92}$Physics Department, University of Cape Town, Cape Town, South Africa
\\
$^{93}$Physics Department, University of Jammu, Jammu, India
\\
$^{94}$Physics Department, University of Rajasthan, Jaipur, India
\\
$^{95}$Physikalisches Institut, Eberhard Karls Universit\"{a}t T\"{u}bingen, T\"{u}bingen, Germany
\\
$^{96}$Physikalisches Institut, Ruprecht-Karls-Universit\"{a}t Heidelberg, Heidelberg, Germany
\\
$^{97}$Physik Department, Technische Universit\"{a}t M\"{u}nchen, Munich, Germany
\\
$^{98}$Purdue University, West Lafayette, Indiana, United States
\\
$^{99}$Pusan National University, Pusan, South Korea
\\
$^{100}$Research Division and ExtreMe Matter Institute EMMI, GSI Helmholtzzentrum f\"ur Schwerionenforschung, Darmstadt, Germany
\\
$^{101}$Rudjer Bo\v{s}kovi\'{c} Institute, Zagreb, Croatia
\\
$^{102}$Russian Federal Nuclear Center (VNIIEF), Sarov, Russia
\\
$^{103}$Saha Institute of Nuclear Physics, Kolkata, India
\\
$^{104}$School of Physics and Astronomy, University of Birmingham, Birmingham, United Kingdom
\\
$^{105}$Secci\'{o}n F\'{\i}sica, Departamento de Ciencias, Pontificia Universidad Cat\'{o}lica del Per\'{u}, Lima, Peru
\\
$^{106}$Sezione INFN, Bari, Italy
\\
$^{107}$Sezione INFN, Bologna, Italy
\\
$^{108}$Sezione INFN, Cagliari, Italy
\\
$^{109}$Sezione INFN, Catania, Italy
\\
$^{110}$Sezione INFN, Padova, Italy
\\
$^{111}$Sezione INFN, Rome, Italy
\\
$^{112}$Sezione INFN, Trieste, Italy
\\
$^{113}$Sezione INFN, Turin, Italy
\\
$^{114}$SSC IHEP of NRC Kurchatov institute, Protvino, Russia
\\
$^{115}$Stefan Meyer Institut f\"{u}r Subatomare Physik (SMI), Vienna, Austria
\\
$^{116}$SUBATECH, Ecole des Mines de Nantes, Universit\'{e} de Nantes, CNRS-IN2P3, Nantes, France
\\
$^{117}$Suranaree University of Technology, Nakhon Ratchasima, Thailand
\\
$^{118}$Technical University of Ko\v{s}ice, Ko\v{s}ice, Slovakia
\\
$^{119}$Technical University of Split FESB, Split, Croatia
\\
$^{120}$The Henryk Niewodniczanski Institute of Nuclear Physics, Polish Academy of Sciences, Cracow, Poland
\\
$^{121}$The University of Texas at Austin, Physics Department, Austin, Texas, United States
\\
$^{122}$Universidad Aut\'{o}noma de Sinaloa, Culiac\'{a}n, Mexico
\\
$^{123}$Universidade de S\~{a}o Paulo (USP), S\~{a}o Paulo, Brazil
\\
$^{124}$Universidade Estadual de Campinas (UNICAMP), Campinas, Brazil
\\
$^{125}$Universidade Federal do ABC, Santo Andre, Brazil
\\
$^{126}$University of Houston, Houston, Texas, United States
\\
$^{127}$University of Jyv\"{a}skyl\"{a}, Jyv\"{a}skyl\"{a}, Finland
\\
$^{128}$University of Liverpool, Liverpool, United Kingdom
\\
$^{129}$University of Tennessee, Knoxville, Tennessee, United States
\\
$^{130}$University of the Witwatersrand, Johannesburg, South Africa
\\
$^{131}$University of Tokyo, Tokyo, Japan
\\
$^{132}$University of Tsukuba, Tsukuba, Japan
\\
$^{133}$University of Zagreb, Zagreb, Croatia
\\
$^{134}$Universit\'{e} de Lyon, Universit\'{e} Lyon 1, CNRS/IN2P3, IPN-Lyon, Villeurbanne, Lyon, France
\\
$^{135}$Universit\`{a} di Brescia, Brescia, Italy
\\
$^{136}$V.~Fock Institute for Physics, St. Petersburg State University, St. Petersburg, Russia
\\
$^{137}$Variable Energy Cyclotron Centre, Kolkata, India
\\
$^{138}$Warsaw University of Technology, Warsaw, Poland
\\
$^{139}$Wayne State University, Detroit, Michigan, United States
\\
$^{140}$Wigner Research Centre for Physics, Hungarian Academy of Sciences, Budapest, Hungary
\\
$^{141}$Yale University, New Haven, Connecticut, United States
\\
$^{142}$Yonsei University, Seoul, South Korea
\\
$^{143}$Zentrum f\"{u}r Technologietransfer und Telekommunikation (ZTT), Fachhochschule Worms, Worms, Germany
\endgroup

\fi
\else 
\iffull
\\

\input{fa_2016-09-09fa_2016-09-09.tex}
\input{references_prl.tex}
\else
\ifbibtex
\bibliographystyle{apsrev4-1}
\bibliography{biblio}{}
\else

\fi
\fi
\fi
\end{document}